# DETERMINING Ω FROM PECULIAR VELOCITIES

AVISHAI DEKEL

*Racah Institute of Physics, The Hebrew University of Jerusalem*

**Abstract**
The large-scale dynamics of matter is inferred from the observed peculiar velocities of galaxies via the POTENT procedure. The smoothed fields of velocity and mass-density fluctuations are recovered from the current data of $\sim 3,000$ galaxies. The cosmological density parameter $\Omega$ can then be constrained in three ways: (a) by comparing the density fields of mass and galaxies, (b) by using the velocity field in voids, and (c) by investigating quasilinear deviations from Gaussian fluctuations. The results indicate a high value of $\Omega \simeq 1$; values in the range 0.1-0.3 are rejected with high confidence.

## 1. Introduction

The raw data are estimated distances $r_i$ and redshifts $z_i$ for a set of objects in directions $\hat{r}_i$ [34] The distances are estimated by the Tully-Fisher method for spirals, and by the analogous $D_n$–$\sigma$ method for ellipticals and S0's. The corresponding radial peculiar velocities are $u_i = cz_i - r_i$ ($H_0$ is set to unity; distances are measured in km s$^{-1}$). Given these sparsely-sampled radial velocities, POTENT first computes a *smoothed* radial-velocity field, $u(\boldsymbol{r})$, in a spherical grid using a tensor window function [8,10]. In the current analysis we use a Gaussian window with radius $\sim 1200$ km s$^{-1}$. Weighting inversely by the local density near each object mimics equal-volume averaging which minimizes the bias due to the nonuniform sampling, and weighting inversely by the distance variance, $\sigma_i^2$, reduces the random effects of the measurement errors.

The velocity field is recovered under the assumption of *potential* flow: $\boldsymbol{v}(\boldsymbol{x}) = -\boldsymbol{\nabla}\Phi(\boldsymbol{x})$ [3]. According to linear gravitational instability theory (GI) any vorticity mode decays in time as the universe expands, and based on Kelvin's circulation theorem the flow remains vorticity-free in the mildely-nonlinear regime as long as the flow is laminar. The velocity potential can therefore be calculated by integrating the radial velocity along radial rays, $\Phi(\boldsymbol{x}) = -\int_0^r u(r',\theta,\phi)dr'$. Differentiating $\Phi$ in the transverse directions then recovers the two missing velocity components.

The underlying mass-density fluctuation field, $\delta(\boldsymbol{x})$, is computed by the approximation [28]

$$\delta_c(\boldsymbol{x}) = \|I - f(\Omega)^{-1} \partial \boldsymbol{v}/\partial \boldsymbol{x}\| - 1 ,  \qquad (1)$$

where the bars denote the Jacobian determinant, $I$ is the unit matrix, and $f(\Omega) \equiv \dot{D}/D \simeq \Omega^{0.6}$ with $D(t)$ the linear growth factor [30]. Eq. 2 is the solution to the continuity equation under the Zel'dovich assumption that particle displacements evolve in a universal rate [35]. This expression still involves only the first partial derivatives in Eulerian space, and it reduces to the familiar $\delta = -f(\Omega)^{-1} \boldsymbol{\nabla} \cdot \boldsymbol{v}$ in the linear regime. It has been found to approximate the true density in N-body simulations with an *rms* error less than 0.1 over the range $-0.8 \leq \delta \leq 4.5$ [26,23].

The results suffer from random and systematic errors due to the distance errors and the nonuniform sampling. The ways we handle these errors are discussed in detail in [8] and in [10], where special attention has been paid to correcting for inhomogeneous Malmquist bias.

Figures 1 and 2 show preliminary maps of the recovered density field, now extending to beyond 8000 km s$^{-1}$ in some regions [10]. The extended data includes the sample of 493 objects [6,22] which was the basis for the original *POTENT*90 analysis [9], and several much bigger spiral samples [7,15,16,24,25,33] – a total of more than 3000 galaxies that we have self-consistently calibrated and carefully put together [12,34].

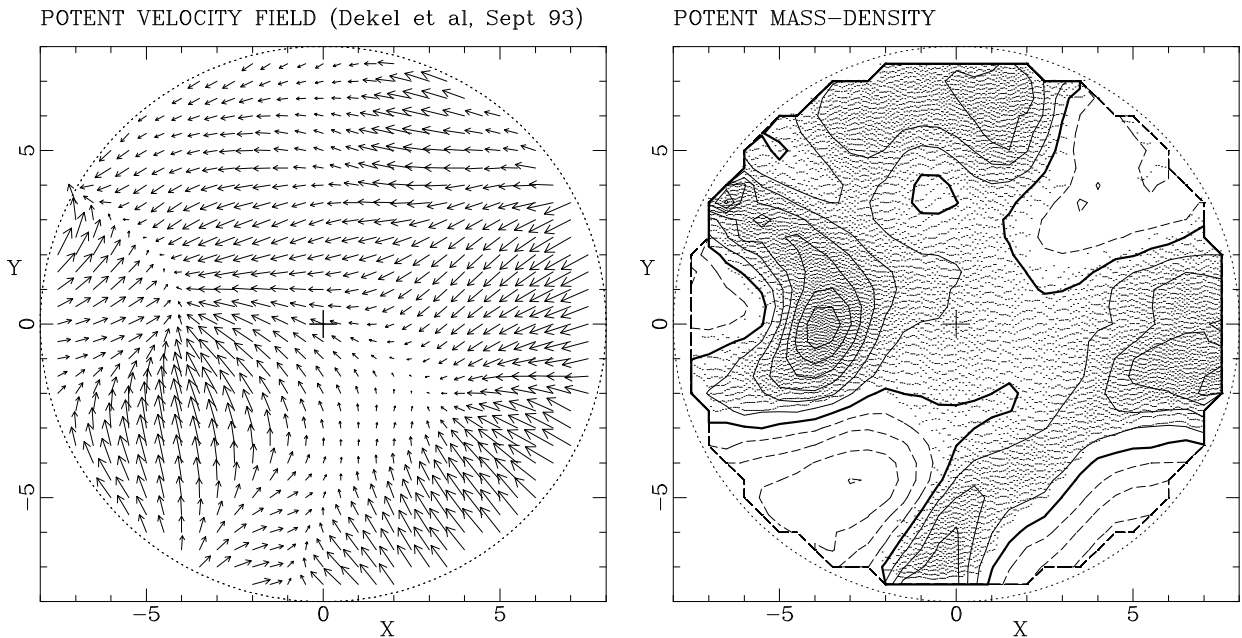

**Figure 1:** Cosmography: the fluctuation fields of velocity and mass-density in the Supergalactic plane as recovered by *POTENT* from the velocities of $\sim 3000$ galaxies with 1200 km s$^{-1}$ smoothing. The vectors shown are projections of the 3D velocity field in the CMB frame. Contour spacing is 0.2 in $\delta$, with the heavy contour marking $\delta = 0$ and dashed contours denoting negative fluctuations. The Local Group is at the origin. The GA is on the left, PP on the right, and Coma is at the top.

The density peak of the Great Attractor (GA) under 1200 km s$^{-1}$ smoothing is of $\delta = 1.2 \pm 0.3$ near the Galactic plane ($X \simeq -4000, Y \simeq 0$). The Perseus-Pisces (PP) peak is at the level of $\delta = 0.7 \pm 0.3$ and it extends towards Aquarius and Cetus near the south Galactic pole (where the "Southern Wall" is seen in redshift surveys). The Coma Great Wall shows up towards the north galactic pole. Two great voids extend from bottom left to top right.

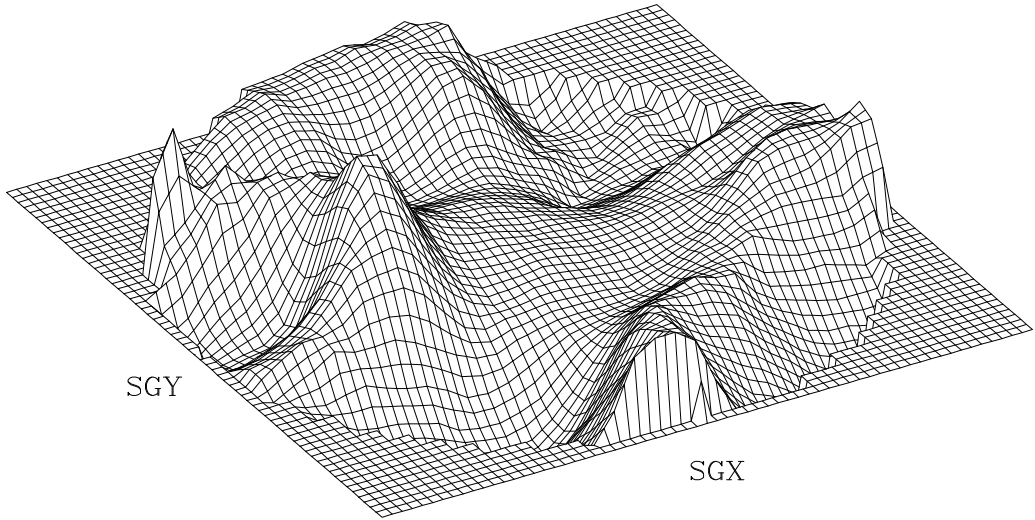

**Figure 2:** Cosmological landscape: mass density fluctuations in the Supergalactic plane as recovered by POTENT from the velocities of $\sim 3000$ galaxies with $1000$ km s$^{-1}$ smoothing. The height of the surface is proportional to $\delta$. The map extends to a distance of $8000$ km s$^{-1}$ about the Local Group. The GA is on the left, PP on the right, and Coma at the top.

Is there a back flow behind the GA? Is PP moving toward the LG or away from it? These effects are detected by the current POTENT analysis still only at the 1.5-sigma level. Recall that there is an important free parameter in the velocity field: the distances are determined only relative to each other and the mean Hubble flow is not known. It has been determined here by minimizing residuals within the volume sampled, but a Hubble-like peculiar velocity error of order 10% is not out of the question. The data thus allow a solution with no GA back flow and with PP moving away from the LG.

A simple and robust statistic is the *bulk velocity* within a sphere of radius $R$ centered on the LG. It is computed by weighted vector averaging of the smoothed peculiar velocities at the grid points. The results suffer from a systematic uncertainty which arises from the non-uniform sampling. We show in Figure 3 two different results, obtained by applying different relative weights to the data; one minimizing the sampling-gradient bias by equal-volume weighting, and the other minimizing the random errors by weighting with the inverse of the distance variances. The difference between the two results reflects the systematic uncertainty. The bulk velocity in a top-hat sphere of radius $6000$ km s$^{-1}$ is thus in the range $270 - 360$ km s$^{-1}$. The additional 1-sigma random error in $|V|$ due to distance errors is typically 15%, as derived from Monte-Carlo noise simulations. This error does not include the random error due to cosmic scatter – the fact that only one sphere has been sampled.

Three arguments support our belief that the peculiar velocities are real, and consistent with being generated by GI: (a) The velocity fields traced separately by spirals and by ellipticals, using different distance indicators, are consistent with each other [21] (b) the COBE detection of $\delta T/T \sim 10^{-5}$ and the bulk velocity of $\sim 350$ km s$^{-1}$ measured in the local neighborhood in a top-hat sphere of radius $6000$ km s$^{-1}$ are consistent with each other [5], and (c) the galaxy density from redshift catalogs and the mass density recovered from observed velocities are consistent with each other (§2; [9]). We thus assume that the velocities are real and generated by the gravity of the mass-density fluctuations, and try to determine $\Omega$.

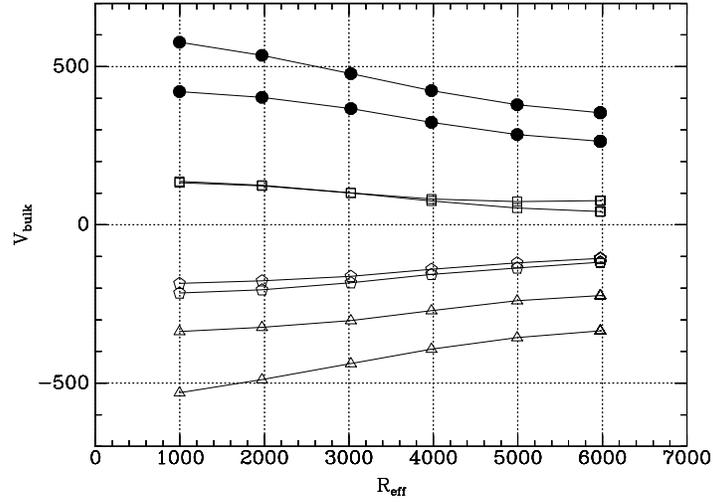

**Figure 3:** The bulk velocity in a top-hat sphere of radius $R_{eff}$ about the LG. Shown are $|V|$ (filled), $V_x$ (triangles), $V_y$ (squares), and $V_z$ (hexagons). The two results shown reflect the systematic uncertainty. The 1-sigma random error due to distance errors is $\simeq 15\%$.

## 2. Galaxies Versus Mass: $\Omega$ and Biasing

The POTENT density $\delta_P$, determined assuming $\Omega = 1$, relates to the true $\delta$ in the linear regime by $\delta_P \propto f(\Omega)\delta$. On the other hand, if we parameterize the relation between the galaxy density fluctuation field, $\delta_G$, and the mass density fluctuations by a universal "biasing" factor, $\delta_G = b\delta$, then we expect a relation of the sort $\delta_P = [f(\Omega)/b]\delta_G$. Given the uncertainties in the two datasets, we can ask whether the POTENT data are consistent with being a noisy version of the galaxy data, and obtain the best-fit value for $f(\Omega)/b$ with associated confidence limits.

The degeneracy of $\Omega$ and $b$ is broken in nonlinear regions, where $\delta(\mathbf{v})$ is no longer simply proportional to $f(\Omega)^{-1}$. The quasilinear analysis based on Eq. 1 allows a first attempt at determining $\Omega$ and $b$ separately.

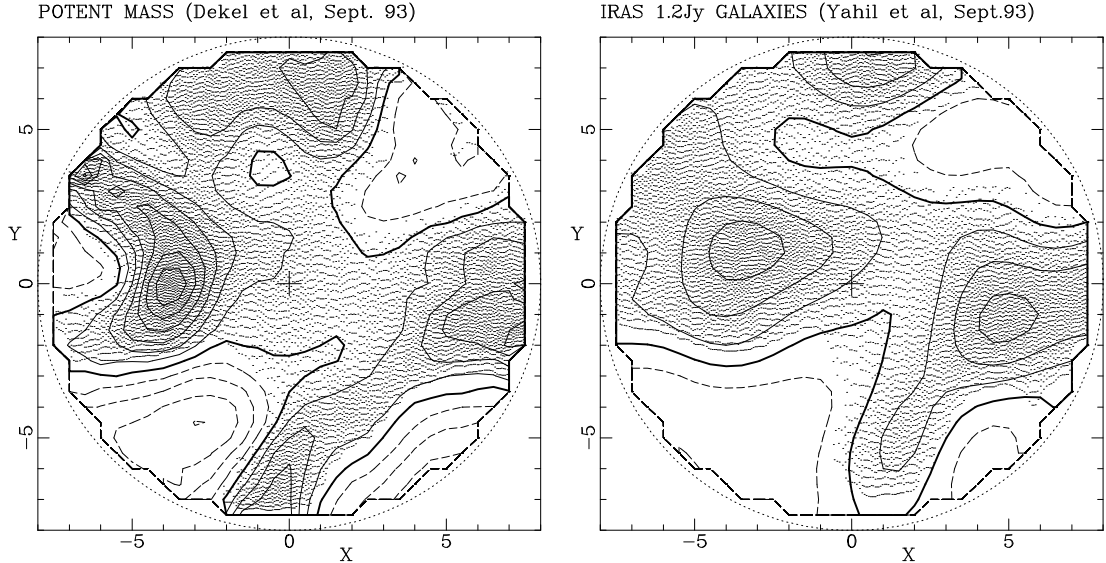

**Figure 4:** Galaxies versus mass density fields in the Supergalactic plane. Contour spacing is 0.2 in $\delta$, with the heavy contour marking $\delta = 0$. The Local Group is at the origin. GA is on the left, PP on the right, and Coma at the top.

Figure 4 compares the density maps in the Supergalactic plane of the POTENT mass density and the galaxy density from a redshift survey of IRAS galaxies, flux limited at 1.2Jy [13], both

Gaussian smoothed at 1200 km s$^{-1}$, assuming $b = 1$. Given the errors, the similarity between the maps is strong. Both feature the general GA phenomenon as a ramp which peaks beyond the Hydra-Centaurus clusters ($X \simeq -4000$) and falls off gradually toward Virgo ($X \simeq -300$, $Y \simeq 1300$) and toward Pavo across the Galactic plane. The fields also agree well in the PP and Cetus superclusters, in the Coma Great Wall region ($X \simeq 0$, $Y \simeq 8000$), and in the great void in between.

These new data have not been subject to a quantitative comparison yet. So far, we have applied an elaborate statistical analysis [9] only to the earlier data of POTENT90 and IRAS 1.9Jy [31], Gaussian smoothed at 1200 km s$^{-1}$. Noise considerations limited that analysis to a volume $\sim (5300$ km s$^{-1})^3$ containing $\sim 12$ independent density samples. Monte-Carlo noise simulations showed that the data are consistent with the hypotheses of linear biasing and GI. The Monte-Carlo simulations were then used to estimate the random errors, to correct for systematic errors in POTENT, and to constrain the parameters via a likelihood analysis. Our robust result is $\Omega^{0.6}/b_I = 1.28^{+0.75}_{-0.59}$ at 95% confidence. Small nonlinear effects allow weaker, separate constraints on $\Omega$ and on $b_I$: if $\Omega = 1$ then $b_I = 0.7^{+0.6}_{-0.2}$, and if $b_I > 0.5$ then $\Omega > 0.46$. Inhomogeneous Malmquist bias could decrease these estimates of $\Omega$, but our 95% confidence limit for $b_I > 0.5$ could be reduced at most to $\Omega > 0.3$. The range of uncertainty is big because the data were limited and because we were extremely careful in our error analysis.

The preliminary results from a similar comparison with the galaxy distribution in the optical catalog [17,18] indicate a similar correlation between light and mass, with $f(\Omega)/b_o \sim 0.7 \pm 0.2$, in agreement with the ratio of $b_o/b_I \simeq 1.5$ obtained by direct comparison, and in agreement with common theoretical ideas. Quantitative comparisons using the extended velocity data and the new IRAS and optical surveys will hopefully be able to provide clearer answers to the fundamental questions concerning the validity of GI, the biasing scheme, and the value of $\Omega$ with a reduced range of uncertainty.

## 3. $\Omega$ from Voids

A diverging flow in an extended low-density region can provide a robust dynamical lower bound on $\Omega$, based on the fact that large outflows are not expected in a low-$\Omega$ universe [11]. The velocities are assumed to be induced by gravity from small initial fluctuations, but no assumptions need to be made regarding their exact Gaussian nature, galaxy biasing, or $\Lambda$.

The derivatives of a diverging velocity field infer a nonlinear approximation to the mass density, $\delta_c$ (Eq. 1), which is an overestimate when the true value of $\Omega$ is assumed []. Analogously to the behavior of the linear approximation, $\delta_0 = -\Omega^{-0.6} \nabla \cdot \boldsymbol{v}$, the $\delta_c$ inferred from a given velocity field is a monotonically increasing function of $\Omega$, and it may become smaller than $-1$ for $\Omega$ values that are too small below the true value. Then, since $\delta \geq -1$ because mass is never negative, the allowed values for $\Omega$ are bounded from below.

Given the observed radial peculiar velocities of galaxies, one can use the POTENT procedure to recover the three-dimensional velocity field, Gaussian smoothed at $\sim 1200$ km s$^{-1}$. One can then derive the inferred mass-density field and the associated *error* field for different values of $\Omega$. Focusing on the deepest density wells, the assumed $\Omega$ should be lowered until $\delta_c$ becomes significantly smaller than $-1$, where the significance is estimated based on the estimated errors. Such low values of $\Omega$ could be ruled out.

The most promising "test case" provided by the current data seems to be a broad diverging region centered near the supergalactic plane at the vicinity of $(X,Y) = (-2500, -4000)$ in km s$^{-1}$ supergalactic coordinates. This region can be roughly identified with the "Sculpter void" of galaxies, clearly seen in the SSRS redshift catalog next to the "Southern Wall".

As can be seen in Figure 5, values of $\Omega \approx 1$ are perfectly consistent with the data, but $\delta_c$

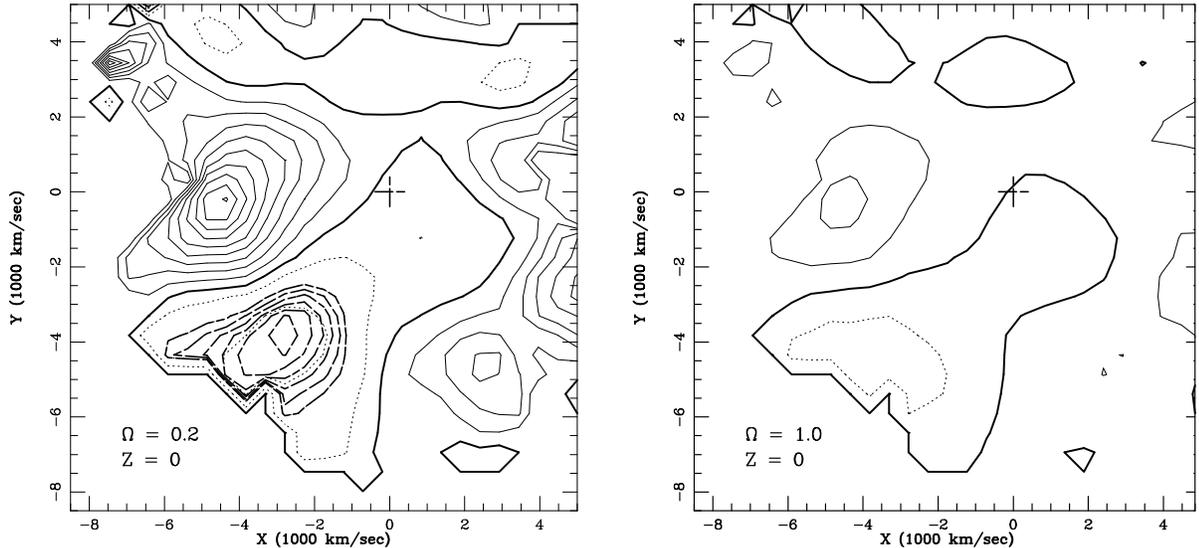

**Figure 5:** Maps of the density-fluctuation field inferred from the observed velocity derivatives, $\delta_c$, in part of the supergalactic plane, for two different values assumed for $\Omega$. The void of interest is confined by the Pavo part of the Great Attractor on the left and the Aquarius extension of the Perseus-Pisces supercluster on the right. The Local Group is marked by a '+'. Contour spacing is 0.5, with the mean, $\delta_c = 0$, marked by a heavy line, $\delta_c > 0$ solid, and $\delta_c < 0$ dotted. The heavy-dashed contours mark the downward deviation of $\delta_c$ from $-1$ in units of the standard deviation $\sigma$, starting from zero (which coincides with $\delta_c = -1$), and decreasing with spacing $-0.5\sigma$. The value $\Omega = 0.2$ is ruled out at the 2.9-sigma level.

becomes smaller than $-1$ in this void already for $\Omega = 0.6$. The confidence by which $\delta_c < -1$ at a given point is expressed in terms of the random error $\sigma$ there. The values $\Omega = 0.4, 0.3$, and 0.2 are ruled out at the 1.6-, 2.4-, and 2.9-sigma levels respectively.

The result reported here is just a preliminary application of the method. The systematic errors have been partially corrected for in the POTENT procedure, but a more specific investigation of the biases affecting the smoothed velocity field in the deepest density wells is required for more reliable results [14].

For the method to be effective we need to find a void that is (a) bigger than the correlation length for its vicinity to represent the universal $\Omega$, (b) deep enough for the lower bound to be strong, (c) nearby enough for the distance errors to be small, and (d) properly sampled to trace the velocity field in its vicinity.

## 4. $\Omega$ from the Probability Distribution Function

A generalization of the method using voids to determine $\Omega$ makes use of the whole one-point probability distribution function (PDF) of the smoothed POTENT density field. Assuming that the initial fluctuations are a random *Gaussian* field, the PDF dvelops a characteristic skewness due to nonlinear effects, which is noticieable early in the quasilinear regime [23]. The skewness of $\delta$, in second-order perturbation theory, is approximated by $\langle\delta^3\rangle/\langle\delta^2\rangle^2 = (34/7 - 3 - n)$, with $n$ the effective power index near the smoothing scale [24]. This result is almost independent of $\Omega$. Since in linear theory $\delta = -f(\Omega)^{-1}\nabla\cdot\boldsymbol{v}$, one can expect the corresponding skewness of $\nabla\cdot\boldsymbol{v}$ to strongly depend on $\Omega$. Indeed, our second-order calculation yields [25]

$$S_3 \equiv \langle(\nabla\cdot\boldsymbol{v})^3\rangle/\langle(\nabla\cdot\boldsymbol{v})^2\rangle^2 = -f(\Omega)^{-1}(26/7 - 3 - n) \ . \qquad (2)$$

Using N-body simulations and $1200 \text{ km s}^{-1}$ smoothing we indeed find $S_3 = 1.8 \pm 0.6$ for $\Omega = 1$ and $S_3 = 4.1 \pm 1.1$ for $\Omega = 0.3$. The quoted error is the standard deviation associated with computing $S_3$ within one single sphere of radius 5000 km s$^{-1}$ in a CDM universe ($H_0 = 75$, $b = 1$).

The value of $S_3$ in the current POTENT velocity field within 5000 km s$^{-1}$ is $0.6 \pm 0.5$, here the error representing distance-measurement errors. The two kind of errors should roughly add in quadrature. Hence, $\Omega = 0.3$ is rejected at the 2.5-sigma level. This bound is similar in strength to the bound obtained using voids, but it relies on the assumption of Gaussian initial fluctuations and it is somewhat sensitive to the exact shape of the power spectrum (mostly through the error estimate).

The method just described still makes use of only a fraction of the data – the PDF – which does not fully specify the field. In fact, the PDF tends to develop a general lognormal shape in the course of quasilinear evolution even in certain cases of non-Gaussian initial fluctuations [26]. A more powerful bound on $\Omega$ can be obtained if one makes use of the detailed POTENT velocity field to recover the *initial PDF* (IPDF), and then use this IPDF to constrain $\Omega$. Nusser and Dekel [27,28] have developed a method for tracing quasilinear fluctuations back in time and determining the IPDF using an Eulerian representation of the Zel'dovich approximation. Then, the key point for determining $\Omega$ is that the IPDF recovered from the present velocities is very sensitive to the value of $\Omega$ assumed in the recovery process (similar in principle but opposite in sign to the $\Omega$ dependence of the present PDF of $\nabla \cdot \boldsymbol{v}$). If we are willing to assume a specific shape for the IPDF based on other considerations, most naturally a Gaussian shape, then we can constrain $\Omega$ by comparing the recovered IPDF with the assumed one.

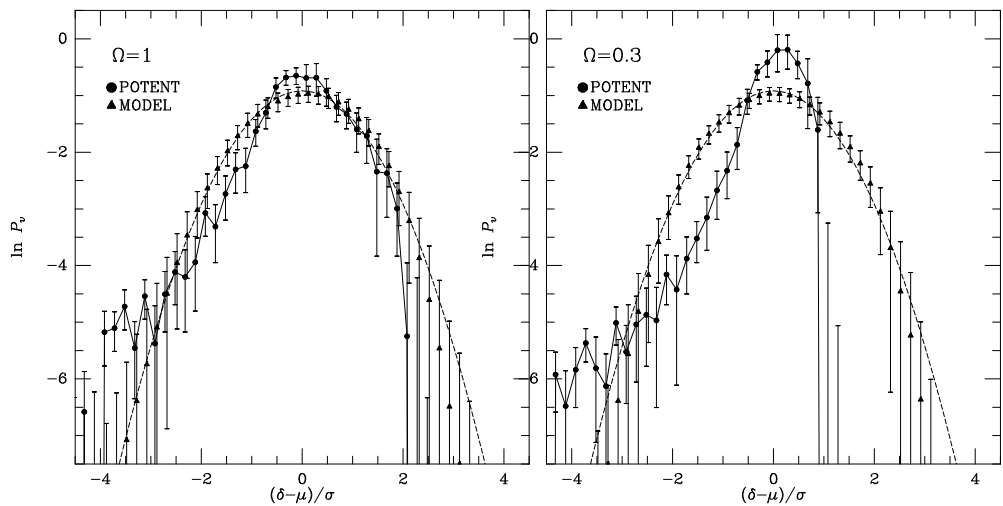

**Figure 6:** The density IPDF recovered from the velocity field provided by POTENT90 from observed velocities (solid) compared with Gaussian (short dash) and with the IPDF recovered from the velocity field of Gaussian CDM simulations (triangles). The assumed $\Omega$ is 1 or 0.3. The simulations are of $\Omega_0 = \Omega$ accordingly.

The velocity field out of POTENT90 within a conservatively selected volume has been fed into the IPDF recovery procedure with $\Omega$ either 1 or 0.3. The recovered IPDF's are shown in Figure 6. The error bars attached to the POTENT IPDF are the Monte-Carlo measurement errors, and the error bars attached to the model IPDF estimate the error due to the limited volume sampled. The total error is roughly a sum in quadrature of the two errors. The immediate impression is that the IPDF recovered with $\Omega = 1$ is marginaly consistent with Gaussian while the one recovered with $\Omega = 0.3$ significantly deviates from a Gaussian. Trying to reject these models we use several different statistics which characterize the IPDF. The standard deviation of each statistic due to measurement and volume errors was evaluated using Monte-Carlo simulations. For example, bin by bin in the IPDF, the largest deviation for $\Omega = 1$ is ∼2-sigma while for $\Omega = 0.3$ there are deviations larger than 4-sigma near $x \equiv (\delta - \mu)/\sigma \approx \pm 1.5$. The standard moments of skewness and kurtosis are poorly determined because they are tail-dominated, but the replacements $\langle x|x|\rangle$ and $\langle |x|\rangle$ do very well: The hypothesis $\Omega = 0.3$ is rejected at the 6.2-sigma level by the former and at 5.3-sigma by the latter, while $\Omega = 1$ is "rejected" at below the 1.6-sigma level by both. Thus, $\Omega \geq 0.3$ is strongly ruled out!

The bounds on $\Omega$ discussed in this section depend on the assumption of Gaussian $IPDF$. But in fact, the $IPDF$ does not have to be assumed: it can be determined from an observed *density* field independently of $\Omega$ [27]. We have recovered the $IPDF$ from the 1.2Jy $IRAS$ density field within 80 h$^{-1}$Mpc of us, assuming $b_I = 1$. The method is not that sensitive to the exact value of $b_I$ in the range $0.5 - 1.5$. The reconstructed $IPDF$ is consistent with Gaussian at the ∼1-sigma level even without taking into account the measurement errors in the density field itself [29].

## 5. Conclusions

The maps of velocity and mass-density fields obtained from extended data of ∼ 3000 galaxies with 1200 km s$^{-1}$ smoothing show new dynamical features such as the Perseus-Pisces ramp extending to Aquarius and Cetus, the Coma Great Wall and the great voids in between. The bulk velocity within a top-hat sphere of radius 6000 km s$^{-1}$ is in the range $270 - 360$ km s$^{-1}$.
Based on the recovered dynamical fields one can put the following constraints on $\Omega$:
• By comparing $POTENT$90 mass density with 1.9Jy $IRAS$ galaxy density: $\Omega^{0.6}/b_I = 1.28^{+0.75}_{-0.59}$ at 95% confidence. The extended data indicate a value near unity with tighter constraints.
• A comparison with optical galaxy density indicates $\Omega^{0.6}/b_o \simeq 0.7$. These results are consistent with $\Omega = 1$, $b_I = 1$ and $b_o = 1.5$ at ∼ 1200 km s$^{-1}$; a much lower value for $\Omega$ would require unrealistic antibiasing for $IRAS$ galaxies.
• The velocity around the great local void requires $\Omega > 0.3$ at the 2.4-sigma level, independent of biasing.
• If we adopt the hypothesis that the initial fluctuations were Gaussian, supported by our finding that the density of $IRAS$ galaxies unambiguously indicates a Gaussian $IPDF$, we find by the skewness of the distribution of $\nabla \cdot \boldsymbol{v}$ that $\Omega > 0.3$ at the 2.5-sigma level.
• Under the same assumption of Gaussian $IPDF$, the $IPDF$ recovered by $POTENT$90 from the observed velocities says that $\Omega > 0.3$ at the 6-sigma level

The velocity data thus indicates a high value for $\Omega$, near unity, with $\Omega \leq 0.3$ strongly ruled out. The uncertainty in this general conclusion is mostly due to inhomogeneous Malmquist bias in the velocity data. Tests of our correction method indicate that this bias cannot affect the resultant $\Omega$ by more than 20%.

**Acknowledgements.** I acknowledge enjoyable collaborations with my colleagues as specified in the text. This research has been supported by the US-Israel Binational Science Foundation and by a Basic Research Grant of the Israeli Academy of Sciences.